\documentclass[fleqn,10pt]{wlscirep}
\usepackage[utf8]{inputenc}
\usepackage[T1]{fontenc}

\title{Self-Adaptive Integrated Photonic Receiver for Turbulence Compensation in Free Space Optical Links}

\author[1]{Andres Ivan Martinez}
\author[1]{Gabriele Cavicchioli}
\author[1]{Seyedmohammad Seyedinnavadeh}
\author[1]{Francesco Zanetto}
\author[1]{Marco Sampietro}
\author[2]{Alessandro D'Acierno}
\author[1]{Francesco Morichetti}
\author[1,*]{Andrea Melloni}

\affil[1]{Dipartimento di Elettronica, Informazione e Bioingegneria, Politecnico di Milano, P.za Leonardo da Vinci, 32, Milano, 20133, Italy}
\affil[2]{Huawei Technologies Italia Srl, Centro Direzionale Milano 2 SNC, 20054 Segrate, Italy}

\affil[*]{andrea.melloni@polimi.it}


\begin{abstract} 
In Free Space Optical (FSO) communication systems, atmospheric turbulence distorts the propagating beams, causing a random fading in the received power. This perturbation can be compensated using a multi-aperture receiver that samples the distorted wavefront on different points and adds the various signals coherently. In this work, we report on an adaptive optical receiver that compensates in real time for scintillation in FSO links. The optical front-end of the receiver is entirely integrated in a silicon photonic chip hosting a 2D Optical Antenna Array and a self-adaptive analog Programmable Optical Processor made of a mesh of tunable Mach-Zehnder interferometers. The photonic chip acts as an adaptive interface to couple turbulent FSO beams to single-mode guided optics, enabling energy and cost-effective operation, scalability to systems with a larger number of apertures, modulation-format and data-protocol transparency, and pluggability with commercial fiber optics transceivers. Experimental results demonstrate the effectiveness of the proposed receiver with optical signals at a data rate of 10 Gbit/s transmitted in indoor FSO links where different turbulent conditions, even stronger than those expected in outdoor links of hundreds of meters, are reproduced.
\end{abstract}

\begin{document}

\flushbottom
\maketitle

\thispagestyle{empty}


\section*{Introduction}


 Free Space Optical (FSO) communication systems operating in the near \cite{Bekkali_2022, Abderrahmen} , mid \cite{Zou_2022, Didier:23}, and even long infrared wavelengths \cite{Joharifar_2023} have received increasing interest in recent years. With respect to radiofrequency wireless, FSO systems are suitable for ultrahigh data rates, offer unlicensed frequency spectrum, and require low power consumption\cite{JAHID_2022}; compared to optical fiber communications, FSO avoids the need for cabling, thus reducing installation time and costs \cite{Khalighi_2014, Yang:17}. These properties set FSO systems as the leading candidates to implement earth-space satellite communications \cite{Popoola:1}, cellular radio network backhaul \cite{Pham}, security \cite{Graceffo}, LiDAR and ISAC (Integrated Sensing And Communication) systems \cite{Poulton, Chow_2024}, and even fiber backup paths \cite{Stotaw}. However, FSO performance is typically impaired and limited by adverse atmospheric phenomena such as air turbulence, temperature gradients, fog and rain, dust, and pollution, which induce beam distortion and wander, in addition to beam deviation, extra attenuation, and link misalignments \cite{Andrews}. Currently, solutions have been proposed to compensate for the optical beam divergence, keep the alignment, and compensate for attenuation, but turbulence effects remain mostly unsolved \cite{Abderrahmen}. 
 
 Atmospheric turbulence introduces a random phase and amplitude perturbation to the propagating optical beams \cite{Zhu:1}. This effect can induce a deep signal fading at the photodetector \cite{Cox}, known as scintillation. In addition to the use of large area photodetectors, having the drawback of an inherent narrow bandwidth and limiting the maximum data rate of the link, in the literature, three different approaches are commonly used to compensate for scintillation: adaptive optics \cite{Paillier:20}, modal diversity receivers \cite{Billault:21}, and spatial diversity receivers \cite{Yang:17}. In the first approach, deformable mirrors adapt to the incoming beam wavefront to correct the phase front, mitigating the scintillation. However, the complexity of their control algorithm limits the bandwidth, making them unsuitable to adequately compensate for the scintillation \cite{billault2022evaluation}. The second approach exploits the fact that a scintillated beam can be seen as the linear combination of various spatial modes \cite{Billault:21}; here, a mode demultiplexer is used to decompose the beam in its fundamental modes, and then each mode is mapped to a single mode fiber (SMF). By coherently combining these signals, it is possible to recover the spatial coherence of the beam.  Although this approach allows a complete beam reconstruction, it requires a mode demultiplexer and SMFs with length difference in the order of a small fraction of the bit duration, i.e., tens of $\mu$m  for 10-100 Gbps signals. Finally, spatial diversity exploits multi-aperture receivers that sample the incoming beam in different positions of the arriving beam front and then recombine the signals coherently. This combination can be done in the electrical domain \cite{Geisler:16}, limited by the processing speed of electronics, or in the optical domain \cite{Yang:17}, with some limits on the scalability. 
 
In this work, we present a multi-aperture self-adaptive FSO receiver integrated into a single silicon photonic chip. The receiver, realized in a commercial silicon photonics platform, includes a 2D Optical Antenna Array (OAA) that samples the incoming beam and is connected with a Programmable Optical Processor (POP) made of a mesh of tunable Mach-Zehnder interferometers (MZIs). The POP self-adapts in real-time and coherently combines the sampled optical signals to a single-mode fiber, thus minimizing the fading of the received signal. Experimental results demonstrate the effectiveness of the approach with intensity-modulated optical signals at a data rate of 10 Gbit/s.

\section*{Results}

\subsection*{Design criteria for multi--aperture integrated photonic FSO receiver }

In this section, we illustrate the approach we used to design the OAA front-end of the FSO-integrated photonic receiver. To this end, we performed numerical simulations to understand the effects of turbulence-induced scintillation on Gaussian beams and extract the spatial coherence parameter that is required to design the multi-aperture OAA and the POP.

\subsubsection*{Turbulence effects on free space Gaussian beams}

\begin{figure}[b!]
\centering\includegraphics[width=15cm]{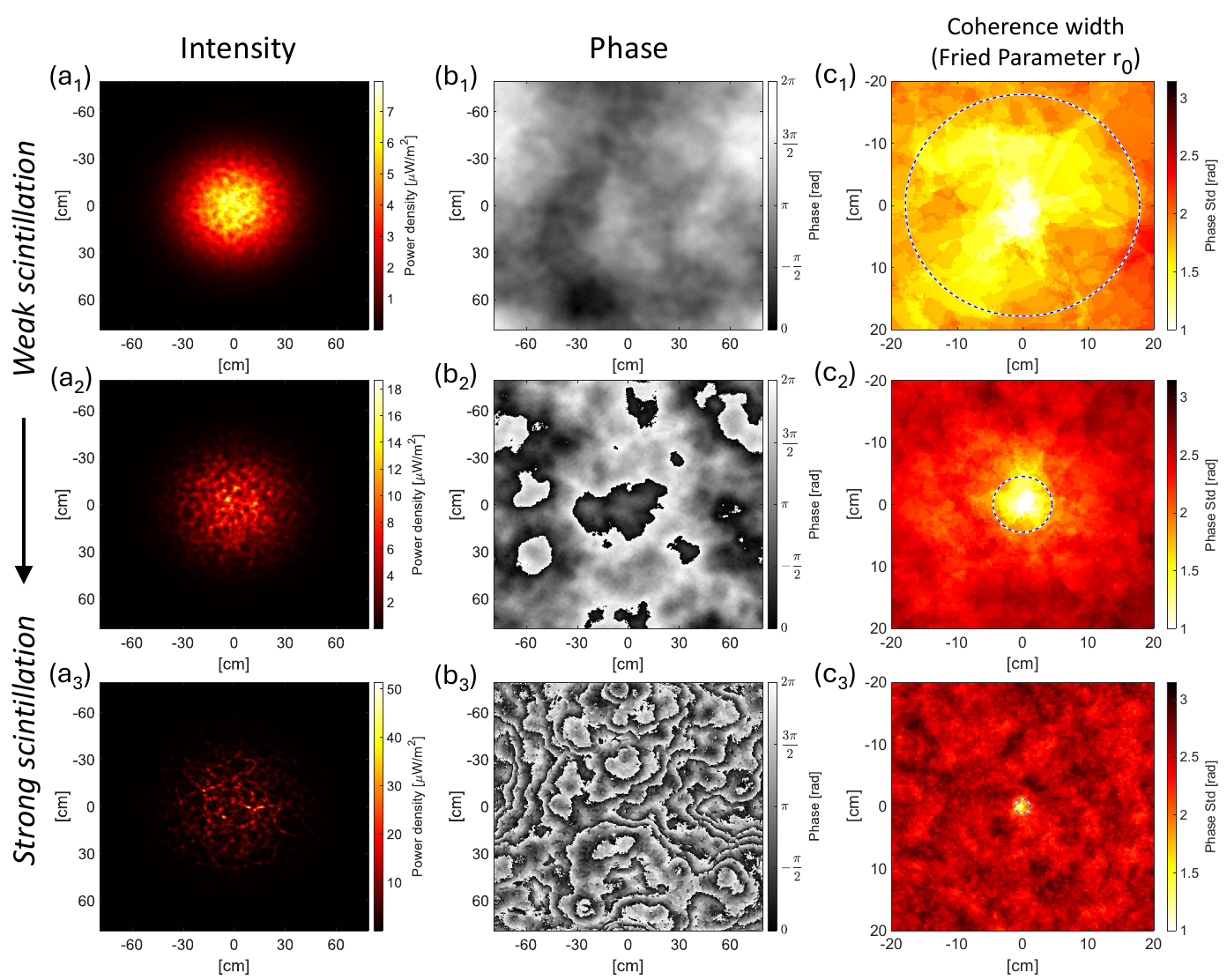}
\caption{\textbf{Effects of turbulence on a Gaussian beam}. Beam with $w_0 = 1$ mm and $\lambda = 1550$ nm propagated for 800 m for low (1$^{\text{st}}$ row), moderate (2$^{\text{nd}}$ row), and strong (3$^{\text{rd}}$ row) turbulence regime. The $1^{\text{st}}$ column corresponds to the beam power density, the $2^{\text{nd}}$ to the phase front, and the $3^{\text{rd}}$ to the spatial coherence. The dashed circles correspond to the theoretical Fried parameter $r_0$.}
\label{fig:SCIeff}
\end{figure}

Figure ~\ref{fig:SCIeff} shows the effects of the turbulence on a Gaussian beam with an initial waist $w_0 = 1$ mm and wavelength $\lambda = 1550$ nm propagated for $L=800$ m under weak ($C_n^2=10^{-15} [\text{m}^{-2/3}]$), moderate ($C_n^2=10^{-14} [\text{m}^{-2/3}]$), and very strong ($C_n^2=10^{-13} [\text{m}^{-2/3}]$) turbulence strength (details on the numerical model are given in the Methods). The beam divergence is 1 mrad full width, and the final diameter of the beam is 80 cm.
The first column (a) shows the spatial power density distribution on the front of the received beam. For weak turbulence (a$_1$), the intensity profile is similar to a pure Gaussian beam. As the turbulence increases, the scintillation increases (a$_2$), and for very strong turbulence (a$_3$), the beam assumes a speckle-like aspect. This characteristic is measured in literature with the Rytov variance \cite{Andrews} that, for the three considered cases, is equal to 0.013, 0.132, and 1.322.

The scintillation also affects the phase front of the beam, as shown in column (b). In (b$_1$), the phase changes smoothly across the wave-front, remaining always within $2\pi$ [rad]. In (b$_2$), the perturbation increases more than $2\pi$ [rad], and therefore, jumps in the phase profile are observed. In (b$_3$), the phase is heavily distorted, presenting frequent phase jumps. The Fried parameter $r_0$, or atmospheric coherence width, defines the diameter of a circular area over which the RMS wavefront aberration equals 1 radian \cite{Fried:66}. The colormaps in Fig.~\ref{fig:SCIeff}c show the spatial coherence of the beam; the dotted circles indicate the theoretically expected coherent area $r_0$, equal to 35, 8.9 and 2.2 cm for the three considered cases. The lower the turbulence (c$_1$), the larger the spatial coherence of the beam $r_0$, which reduces rapidly with the accumulated scintillation. These amplitude and phase patterns of the received beam evolve with the time scale of atmospheric turbulence, which is in the order of a few hundred Hz \cite{Cox}.

\begin{figure}[b!]
\centering\includegraphics[width=17cm]{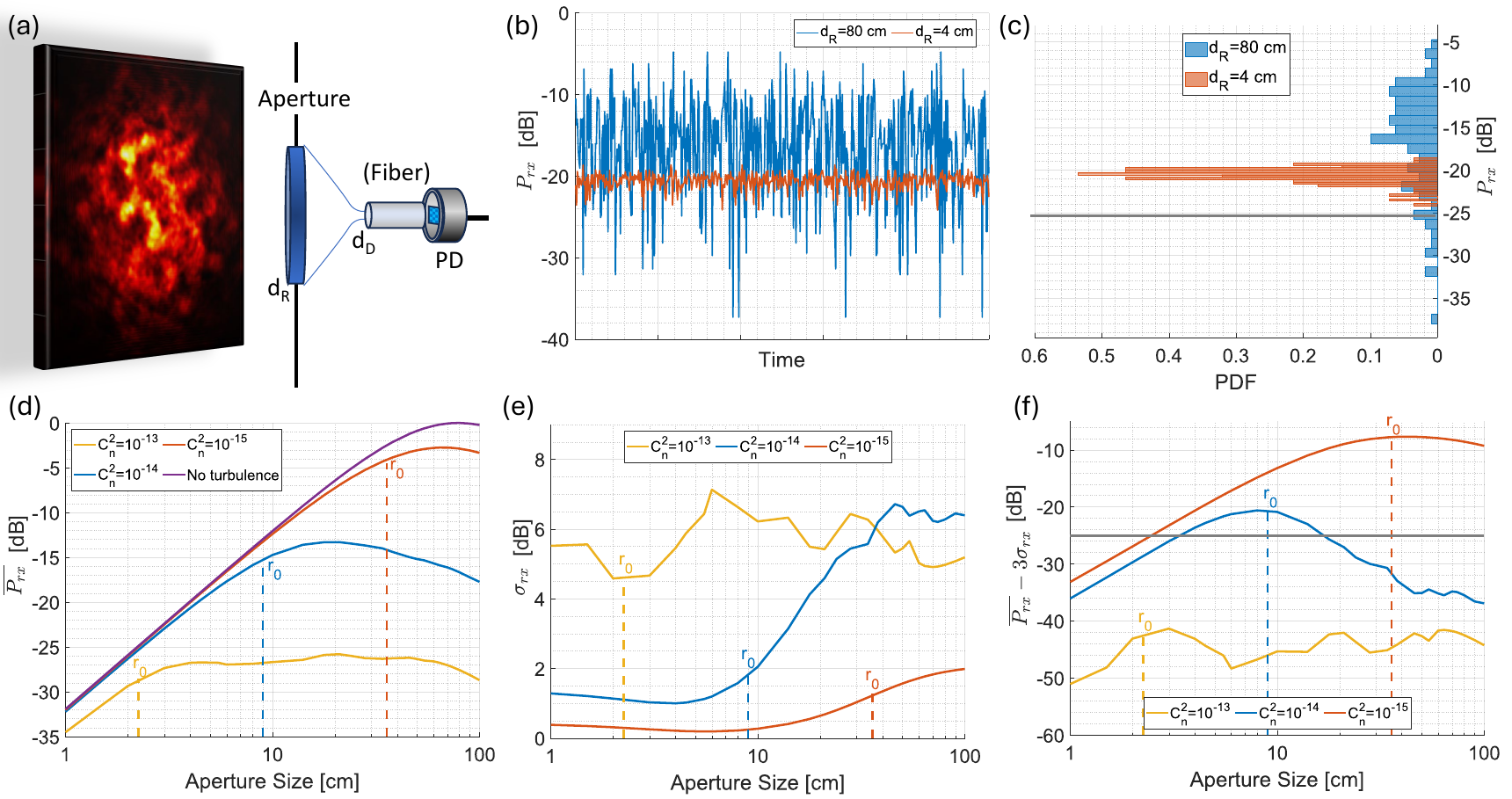}
\caption{\textbf{Single-aperture receiver}. \textbf{a}, Scheme of the single-aperture receiver that focuses a portion of the incoming beam on a single-mode fiber or PD. \textbf{b-c}, Simulated time trace (\textbf{b}) and PDF (\textbf{c}) of the optical power coupled to the PD for a large and small aperture $d_R$ with respect to $r_0$. \textbf{d-f}, Simulated mean received power (\textbf{d}), standard deviation (\textbf{e}), and three-sigma limit (\textbf{f}) of a single aperture receiver with a single-mode detector as a function of the aperture diameter $d_R$ for $C_n^2 =10^{-13}$ (yellow), $10^{-14}$ (blue), $10^{-15}$ (red), and without turbulence (purple). $\overline{P_{rx}}$ and $\sigma_{rx}$ have been calculated as the average of 100 simulations.}
\label{fig:single aperture}
\end{figure}

\subsubsection*{Single-aperture receiver}
A single aperture receiver is considered here, and the main limitations are discussed, setting the motivations for multi-aperture receivers in FSO applications. The main purpose of the receiver is to couple as much power as possible from the incoming beam and provide an output power not affected by time-varying fading due to beam scintillation.  
We consider the system sketched in Fig.~\ref{fig:single aperture}a as a reference single-aperture receiver: a generic input optical aperture with a diameter $d_R$ is coupled to some optics (lens or telescope), which focuses the collected portion of the incoming beam to an output aperture with spot size $d_D$. Such output aperture can be a small-area photodetector (PD), an optical fiber, or a waveguide. Ideally, $d_R$ should be as large as possible. At the same time, $d_D$ has to be small in the case of single-mode fibers or high-speed integrated PDs. The smaller the active area, the lower the PD’s junction capacitance and, in turn, the resistor-capacitor (RC) time constant \cite{Alkhazragi:21}. In contrast, the distorted intensity and phase of the beam are inherently associated with a multimode nature, thus reducing the minimum achievable size of the focused beam on $d_D$. In other words, the loss of spatial coherence of the scintillated beam limits the size at which it can be concentrated using an optical system, that is, the ratio $d_R/d_D$ is limited by the étandue constraint \cite{Einhaus}. The PD size can be set as a trade-off between coupling efficiency and bandwidth. Still, in any case, the performance of the system in the presence of scintillated beams is severely limited, making it difficult to achieve data rates above 1 Gbit/s over distances of a few hundred meters.


To better understand the effect of beam scintillation on the optical power $P_{rx}$ detected by the single-mode receiver, Fig.~\ref{fig:single aperture}b shows the time traces of $P_{rx}$ for two apertures larger (80 cm, blue) and smaller (4 cm, red) than $r_0$ (8.9 cm). Throughout the paper, the effects of a misalignment of the receiver are neglected. The average detected power with the large aperture is higher, but it suffers from an intense fading caused by the multimodality of the beam. In contrast, with the small aperture, $P_{rx}$ is much less affected by the scintillation, even if its mean value is lower. Fig.~\ref{fig:single aperture}c shows the probability density functions of $P_{rx}$ for the two considered apertures.

Figure~\ref{fig:single aperture}d shows a simulation of the relative mean received power $\overline{P_{rx}}$ as a function of the input aperture size $d_R$ for no turbulence (purple), weak $C_n^2=10^{-15} [\text{m}^{-2/3}]$ (red), moderate $C_n^2=10^{-14} [\text{m}^{-2/3}]$ (blue), and strong $C_n^2=10^{-13} [\text{m}^{-2/3}]$ (yellow) turbulence strength. The corresponding values of $r_0$ are marked as a reference. Figure~\ref{fig:single aperture}e shows the corresponding standard deviation $\sigma_{rx}$ of $P_{rx}$, i.e., the fading. Generally, the stronger the turbulence, the lower the mean received power and the higher the fading. As long as $d_R<r_0$, the received power increases with the size of the aperture, is almost independent of the turbulence strength, and has a $\sigma_{rx}$ that slightly reduces with $d_R$. The received power reaches a maximum for $d_R$ slightly higher than $r_0$, meaning that there is no significant gain in increasing the aperture size further. Instead, the standard deviation $\sigma_{rx}$ increases rapidly for apertures approximately larger than $r_0/3$ and above $r_0$ reaches values that can be very detrimental for high data rate optical communication systems. As a result, the scintillation limits the maximum diameter of the aperture. See Supplementary Information sec.~\ref{sec: sup single aper} for details.

As in any communication system, also in FSO it is important to guarantee a low probability that the received power drops below a given sensitivity threshold, say -25 dB, to avoid out-of-service of the link. Figure~\ref{fig:single aperture}f shows the quantity $\overline{P_{rx}}-3 \sigma_{rx}$. It is evident that apertures comparable with $r_0$, despite a lower mean $\overline{P_{rx}}$, guarantee the largest probability to remain above the sensitivity threshold, at least for moderate turbulence, and are therefore the best choice. The aperture size should be selected for the maximum turbulence $C_n^2$ that has to be compensated because $\overline{P_{rx}}-3 \sigma_{rx}$ improves with the reduction of the turbulence.

\subsubsection*{Multi-aperture receiver}

\begin{figure}[t!]
\centering\includegraphics[width=17cm]{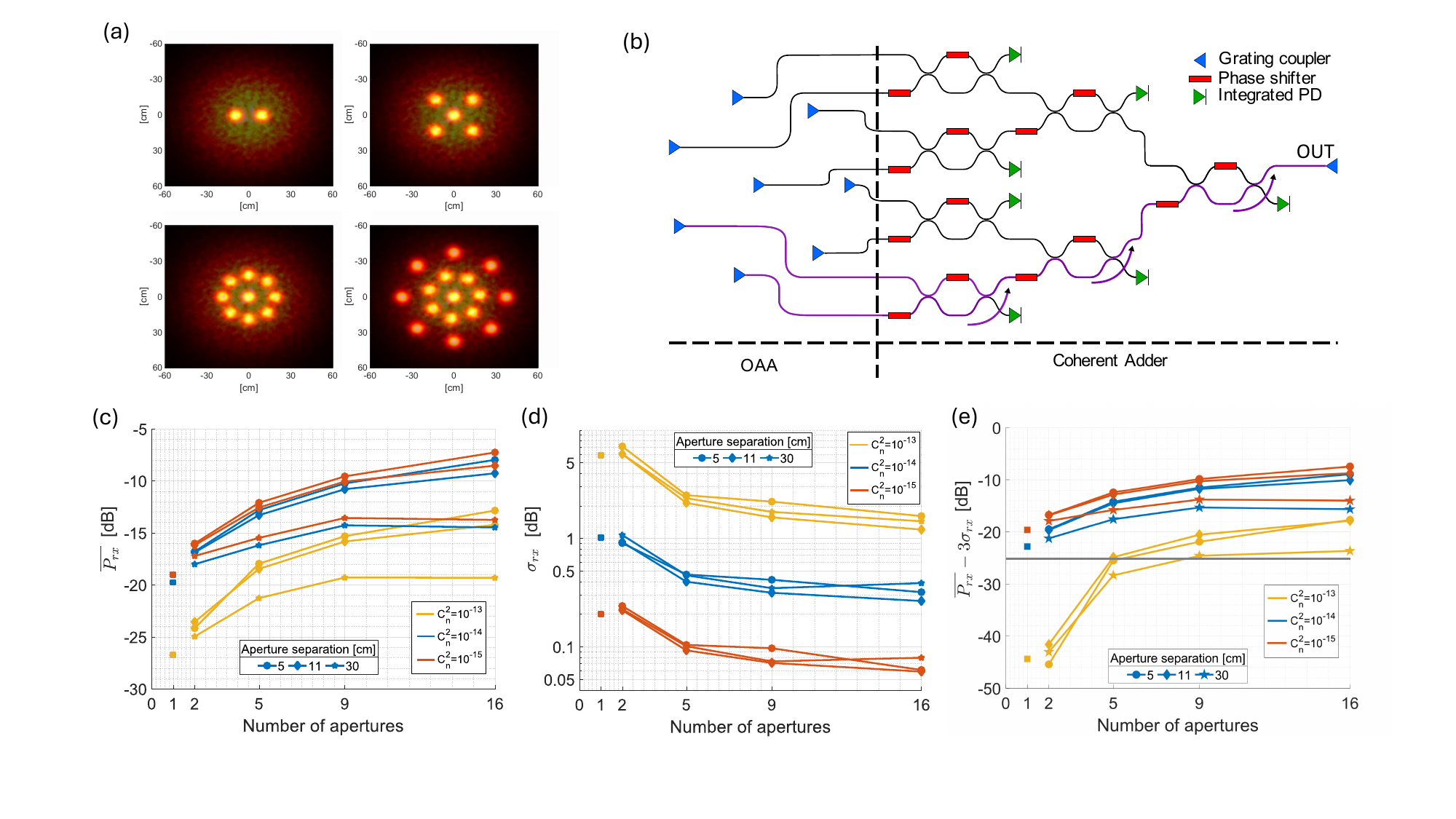}
\caption{\textbf{Multi-aperture receiver structure}. \textbf{a}, Optical Antenna Array distribution for $M = 2$, 5, 9, 16. \textbf{b}, Schematic of the Programmable Optical Processor connected to the OAA. \textbf{c-e},  Mean received power (\textbf{c}), standard deviation (\textbf{d}), and three-sigma limit (\textbf{e}) as a function of the number of apertures for different separation distance $d_S=5$, 11 and 30 cm, and the three considered turbulences $C_n^2= 10^{-13}$, $10^{-14}$, and $10^{-15}$ $[\text{m}^{-2/3}]$. $d_R = 4.5$ cm. $\overline{P_{rx}}$ is calculated by summing coherently the contributions of the $M$ apertures. $\overline{P_{rx}}$ and $\sigma_{rx}$ have been calculated as the average of 100 simulations.}
\label{fig:multi aperture}
\end{figure}

A small aperture guarantees a received power that is almost scintillation insensitive, and it is suitable for high bit rates but has the drawback of coupling just a small portion of the incoming power. A possible solution to achieve higher received power, with minor fading and large bandwidth, is to sample the received beam using several small apertures (see  Fig. \ref{fig:multi aperture}a) and then combine the samples by adding their power. In our approach, the received beam is sampled by an Optical Antenna Array that is integrated directly into the photonic chip. Pairs of samples are coherently combined using balanced tuneable MZIs, which are cascaded to form a binary-tree mesh as shown in Fig. \ref{fig:multi aperture}b. The MZI mesh implements a Programmable Optical Processor (POP) that can be adapted and reconfigured in real time to follow the time-varying scintillation of the received beam. To this end, each MZI has an analog electronic feedback loop that controls the two phase shifters based on the power read by the on-chip integrated slow photodetectors \cite{Zanetto:1}. Different topologies of MZI meshes have been proposed in the literature, with a single output that can be connected to an optical fiber, a PD, or a standard receiver. Remarkably, the POP configuration and tracking time have to match the bandwidth of the scintillation and not that of the signal data rate. A comprehensive description of the working principle of the POP is reported in \cite{Miller:13, Milanizadeh2022, SeyedinNavadeh:24}. The circuit schematic is shown in Fig. \ref{fig:multi aperture}b with an 8 apertures OAA that samples the incoming beam and the POP that sums the samples coherently. This architecture is easily scalable to a more significant number of optical antennas without increasing the control complexity.

The main design parameters of the multi-aperture receiver are the number of apertures $M$ of the OAA, the size of the apertures $d_R$, and the distance $d_S$ between them. In the following numerical analysis, the size $d_D$ is assumed to be suitable for the coupling with single-mode waveguides, and the POP is assumed to work properly, i.e., the output power is the sum of the power received by all the apertures. To determine the optimal values for $M$ and $d_S$, we fixed $d_R = 4.5$ cm (from Fig.~\ref{fig:single aperture}) as a good compromise between a high $P_{rx}$ and low $\sigma_{rx}$ for $C_n^2 \geq 10^{-13}$ (see Supplementary Information sec.~\ref{sec: sup multi aper} for details).  

Figure~\ref{fig:multi aperture}c shows the mean power $\overline{P_{rx}}$ (overall power received by the OAA) vs. the number of apertures $M$ for the three values of $C_n^2$ and three separations of adjacent apertures $d_S= 5$, 11, and 30 cm. The arrangement of the apertures of the OAA is shown in Fig.~\ref{fig:multi aperture}a. As expected, the power $\overline{P_{rx}}$ increases with the number of apertures and saturates when the OAA size is comparable to or larger than the beam width.  $\overline{P_{rx}}$ is weakly dependent on the separation distance unless the size of the OAA is larger than that of the beam. The great advantage of using a multi-aperture receiver is that the coherent combination of the $M$ signals reduces the $\sigma_{rx}$ of the output power, especially if the contributions are uncorrelated, that is, when $d_S>r_0$ because the single fluctuations tend to compensate. Instead, there is no diversity gain when the apertures are within the same coherence area ($d_S < r_0$). In any case, the reduction of the fading improves with $M$, as shown in Fig.~\ref{fig:multi aperture}d. 

As a result of this numerical analysis, the OAA has to cover the most significant portion of the beam with a sufficiently large number of apertures with individual size and mutual distance comparable with the smaller $r_0$ to be compensated. Fig.~\ref{fig:multi aperture}e shows that the quantity $\overline{P_{rx}}-3 \sigma_{rx}$ increases with these choices and guarantees the largest probability of remaining above the sensitivity threshold (-25 dB), indicated by the gray horizontal line. It is worth noting that a multi-aperture receiver with $M \geq 9$ satisfies the minimum sensitivity request of -25 dB also with the strongest considered turbulence, with a significant advantage with respect to the single aperture results shown in Fig.~\ref{fig:single aperture}f.

\subsection*{Integrated FSO photonic receiver}
\begin{figure}t!]
\centering\includegraphics[width=14cm]{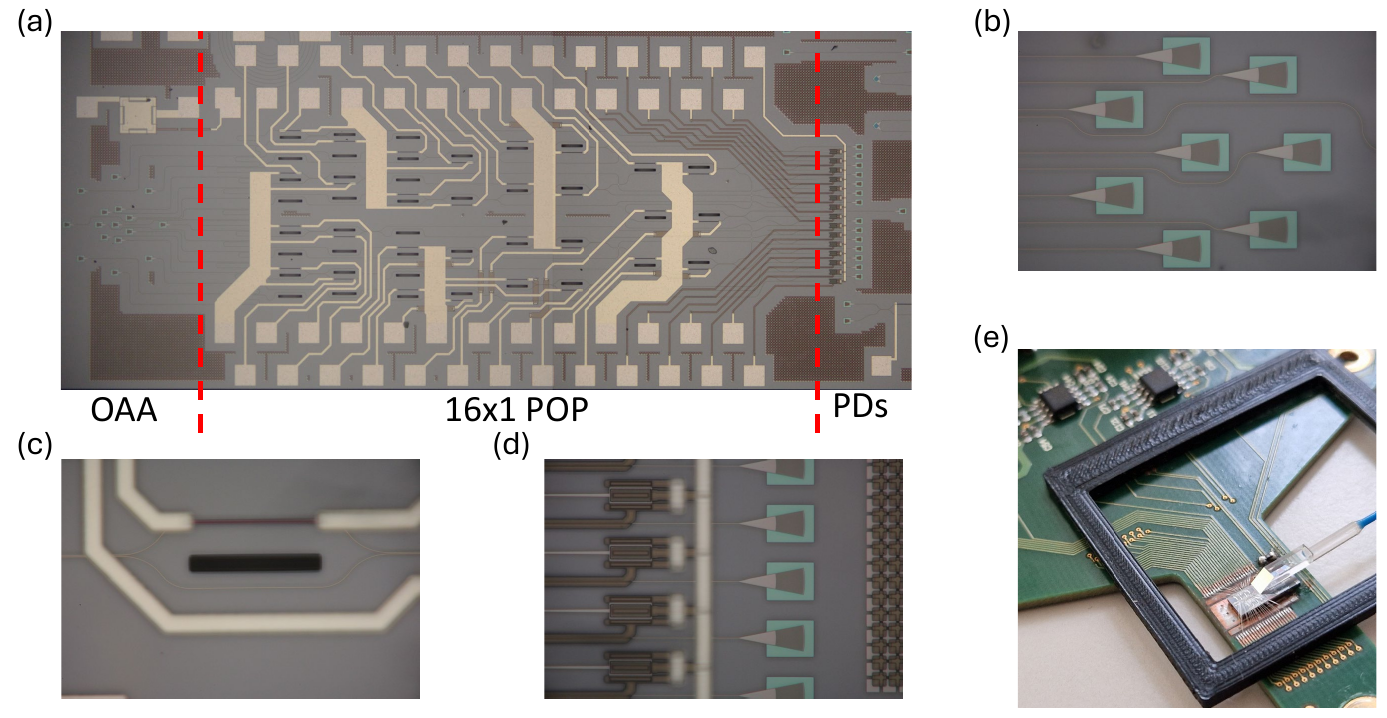}
\caption{\textbf{Integrated silicon photonics FSO receiver}. \textbf{a-e}, Photograph of the fabricated PIC (\textbf{a}), the inner ring of the OAA (\textbf{b}), a single Mach-Zehnder interferometer (\textbf{c}), integrated Ge photodiodes for feedback control (\textbf{d}), and the chip assembled on a PCB with control electronic and fiber transposer (\textbf{e}).}
\label{fig:PIC}
\end{figure}

An integrated photonic FSO receiver has been realized according to the design criteria described in the previous section \cite{SeyedinNavadeh:24}. The receiver operates in the C-band, and it has been fabricated on a standard 220-nm silicon photonic platform by the commercial foundry AMF in an active MPW run. 
The total footprint is 1.3 mm $\times$ 3.2 mm (including the electrical pads) and comprises three sections: a 2D OAA, a POP, and an array of Ge PDs. A photograph of the chip is shown in Fig.~\ref{fig:PIC}a. The OAA is composed of 16 surface grating couplers (GCs). Each GC is 48 $\mu$m-long and 23 $\mu$m-wide, with a 24 $\mu$m-long taper. The radiated field of each GC has an elevation angle of 12°, an azimuth angle of 0°, and a full-width divergence of 5.6° $\times$ 9.8°. The GCs of the OAA are distributed as follows: a central GC, an inner ring of 60 $\mu$m radius with 7 GCs, and an outer ring of 180 $\mu$m radius with 8 GCs. A photograph of the inner ring is shown in Fig.~\ref{fig:PIC}b. The distances between the GCs have been optimized to match the coherence radius of the beam in the maximum turbulence conditions ($d_S \approx r_0$). Note that here, the considered value for $r_0$ is after the demagnification of the collimating optics used to couple the FSO beam (see "Experimental results").

The POP has a 16×1 binary-tree mesh architecture with thermally tuneable MZIs (see Fig.~\ref{fig:PIC}a,c). Each MZI has two actuators (TiN thermal phase shifters), a trench between the arms to reduce the thermal crosstalk, and an integrated Ge PD (see Fig.~\ref{fig:PIC}d and Fig. \ref{fig:multi aperture}b), placed in one of the output ports, that provides a feedback signal to the control loop. The POP is connected through wire bonding to a custom electronic board that minimizes the signal read by the PDs (see Fig.~\ref{fig:PIC}e). In this way, the two optical signals at the input of each MZI are coherently recombined and routed towards a common waveguide (not connected to a PD) \cite{Miller:13}. The control system uses a dithering technique applied to the phase shifters and a lock-in readout of the integrated PDs, which allows the control of the MZIs with no ambiguity on the direction of tuning (see Methods) \cite{Zanetto:1}. In this way, every single actuator can be controlled independently, and all the MZIs can be tuned in parallel, speeding up the setting up of the POP and the tracking of incoming beam scintillation. Finally, the coherently recombined signal is coupled using a single-mode fiber array glued on the chip. 


\subsection*{Experimental results}
To validate the mitigation of turbulence-induced scintillation performed by the integrated photonic receiver, we built an indoor 
experimental setup emulating realistic conditions of an outdoor FSO link. 

\begin{figure}[b!]
\centering\includegraphics[width=17cm]{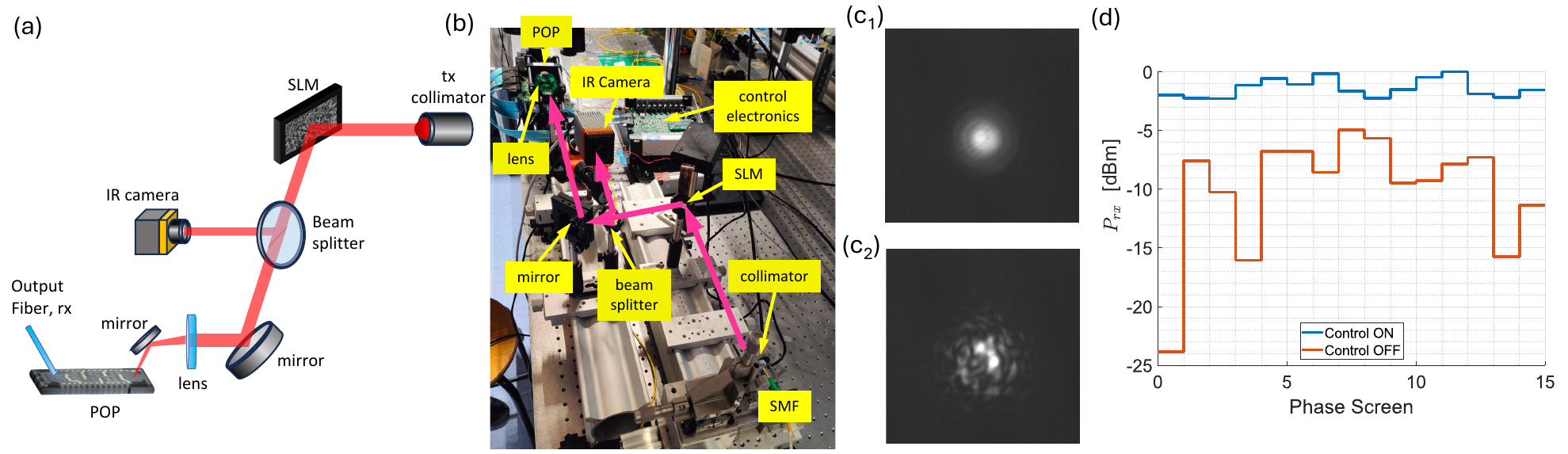}
\caption{\textbf{Static turbulence-induced scintillation compensation}. \textbf{a-b}, Schematic (\textbf{a}) and photo of the experimental setup (\textbf{b}) used for the emulation of turbulence-induced scintillation on FSO beams. \textbf{c}, Image of the optical beam (\textbf{c$_1$}) without and (\textbf{c$_2$}) with the distortion introduced by the SLM phase screen. \textbf{d}, Optical power of the FSO beam at the output port of the integrated receiver for different phase screens when the adaptive control is idle (orange) and active (blue).}
\label{fig:SLM setup}
\end{figure}

As a first demonstration, we used a spatial light modulator (SLM) to introduce controllable static perturbations of the FSO beam. The schematic and the photo of the setup are shown in Fig.~\ref{fig:SLM setup}a,b, respectively (details on the setup are given in the Methods). Panel (c$_1$) shows the beam shape acquired with a near-infrared (IR) camera when the SLM acts as a mirror, while (c$_2$) shows how the phase screen generated by the SLM distorts the beam. Figure~\ref{fig:SLM setup}d shows the optical power received at the output port of the POP for different PSs introduced by the SLM (the PSs used are shown in Supplementary Fig.~\ref{fig:sup SLM ps}). When the control is idle (orange), the average received power is -10.1 dBm with a standard deviation $\sigma_{rx}$ = 5.0 dB. In this case, the signals sampled by the 16 GCs of the OPA combine randomly (constructively or destructively) depending on the local phase in the sampling points of the beam front of the incoming scintillated beam. On the other hand, when the control is active (blue), the POP compensates for the relative phase differences between the samples, maximizing the output power. In this way, the average power increases by 8.7 dB (to -1.4 dBm), and the fading standard deviation reduces to 0.8 dB. The residual fading in the output power is due to the wander effect that deviates the beam's trajectory and cannot be compensated by the adaptive receiver.

\begin{figure}[t!]
\centering\includegraphics[width=17cm]{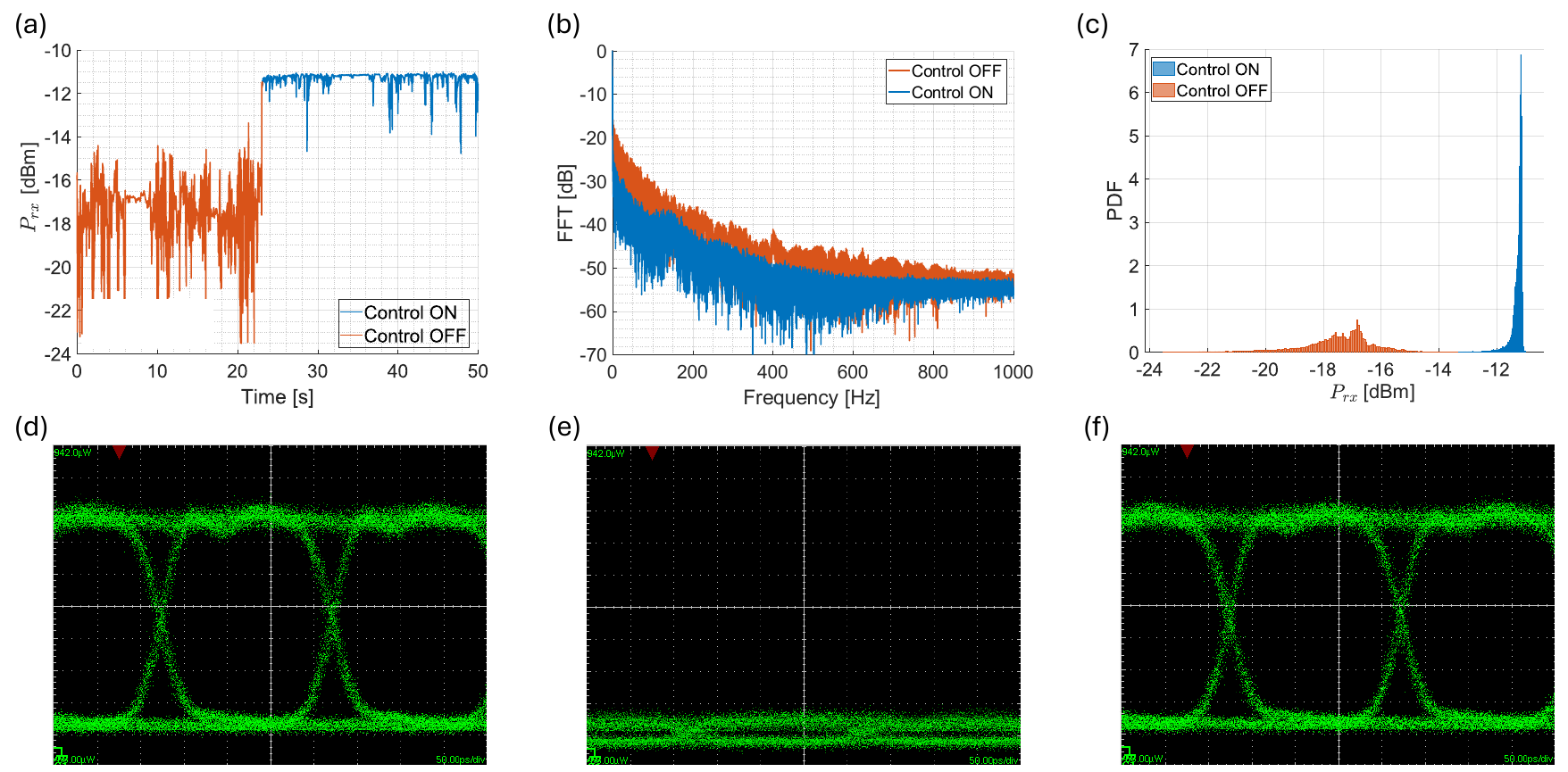}
\caption{\textbf{Real-time turbulence compensation with 5 Gbit/s signal}. \textbf{a-c}, Received optical power $P_{rx}$ (\textbf{a}), power spectral density (\textbf{b}), and probability density function (\textbf{c}) of the received optical power when the POP is in its native state (orange) and when the adaptive control is active (blue). \textbf{d-f}, Eye diagrams of the received 5 Gbit/s OOK signal in the absence (\textbf{d}) and in the presence of turbulence when the adaptive control is off (\textbf{e}) and active (\textbf{f}).}
\label{fig:ctrl OFF}
\end{figure}

To test the receiver in time-varying conditions and verify the speed of the adaptive control, a dynamic test was performed by replacing the SLM with a heat gun (see Methods). Figure~\ref{fig:ctrl OFF}a shows the received optical power $P_{rx}$ acquired for 50 sec in two different conditions. When the POP is off (orange), the main paths across the MZI mesh are the ones passing through the cross ports of the MZIs. Therefore, the $P_{rx}$ is not optimized, presenting a low mean power $\overline{P_{rx}}=-17.5$ dBm and a large standard deviation $\sigma_{rx}=1.3$ dB. When the control loop is active (blue), the average received power at the output port is -11.3 dBm with a standard deviation of only 0.3 dB. The residual minor fading is due to the fill factor of the OAA. In fact, the geometrical loss caused by the spacing between the GCs depends on the pattern of the incoming distorted beam, and the total power sampled by the OAA depends on it. This issue can be mitigated by increasing the fill factor of the OAA, for instance by integrating arrays of microlenses on top of the GCs (see Discussion). 

The spectral power density of the output signal is shown in Fig.~\ref{fig:ctrl OFF}b. The components associated with the artificial turbulence induced by the heat gun (orange) extend up to a frequency range of about 500 Hz, which is in line with the expected spectrum of real atmospheric turbulence (300 Hz) \cite{Cox}. Results show that the real-time adaptation of the POP reduces these components by more than 7 dB up to 400 Hz (blue), which is sufficient to compensate for natural turbulence. Despite the residual fading of the received power with the control ON, the probability density function (PDF) of $P_{rx}$ (see Fig.~\ref{fig:ctrl OFF}c) shows a narrow distribution with most of the power concentrated around -11.3 dBm. Instead, when the control loop is not operative, the $P_{rx}$ PDF has its maximum around -17 dBm and spreads through several dB (from -24 to -14.5 dBm). Fig.~\ref{fig:ctrl OFF}d shows the received eye diagram for a 5 Gbit/s signal without turbulence. In the presence of turbulence with $C_n^2 \approx 10^{-15}$, Fig.~\ref{fig:ctrl OFF}e shows the closed-eye diagram when the POP is in its native state. Instead, when the control loop is active (Fig.~\ref{fig:ctrl OFF}f), the eye remains open in error-free conditions, proving the effective mitigation of scintillation effects.

\begin{figure}[t!]
\centering\includegraphics[width=17cm]{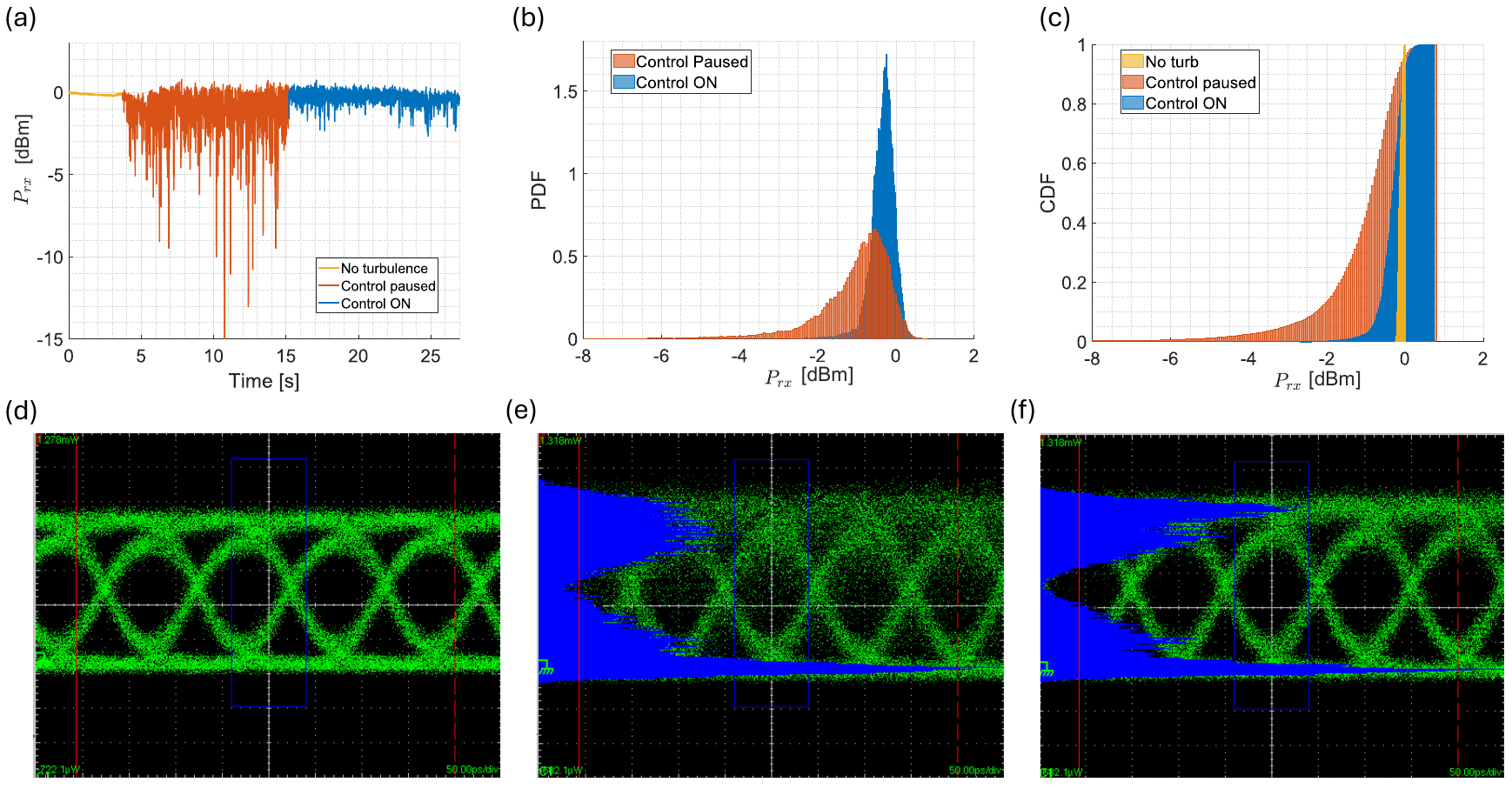}
\caption{\textbf{Real-time turbulence compensation with 10 Gbit/s signal}. \textbf{a-c}, Optical power (\textbf{a}), probability density function (\textbf{b}), and cumulative density function (\textbf{c}) of the received optical signal when the control is active (blue), idle (orange), and in the absence of turbulence (yellow). \textbf{d-f} Corresponding eye diagrams of the received 10 Gbit/s OOK.}
\label{fig:ctrl pause}
\end{figure}

Figure~\ref{fig:ctrl pause} displays the results for a 10 Gbit/s signal. In panel (a), the POP is initially optimized to maximize the coupled power for an incoming beam not affected by turbulence (yellow). Then, the POP is kept in such a fixed state (meaning that all the phase shifters are held in that state), but the heat gun is switched on (orange); finally, the adaptive control is enabled (blue) to track the evolution of the incoming beam distorted by the effect of the heat gun. In this experiment, since the POP was already optimized to maximize the coupling with the undistorted beam, the mean $P_{rx}$ in the idle case with turbulence (-1.1 dBm) is close to that of the active case (-0.3 dBm). Yet, $\sigma_{rx}$ is clearly different in the two cases, which show almost the same values as the ones observed in Figure~\ref{fig:ctrl OFF}a, being 1.1dB for the idle state and 0.3dB for the compensated one, as shown in Fig.~\ref{fig:ctrl pause}a. The cumulative density function CDF is reported in panel (c), showing that the outage probability is always higher for the idle case. The CDF in the absence of turbulence is also reported in yellow as a reference.
Finally, the eye diagrams for the reference perturbed and compensated cases are displayed in Fig.~\ref{fig:ctrl pause}d-f, respectively. The eye in the presence of scintillation is closed and shows a time-varying fluctuation (e). Instead, the eye with the POP that cancels the turbulence is well open and indistinguishable from the reference one.

\section*{Discussion}
In conclusion, we demonstrated the effectiveness of an integrated adaptive receiver based on a 2D optical antenna array and a programmable optical processor to compensate for the scintillation effects in a turbulent FSO link. Design rules regarding the number, size, and distance between the optical antennas of the OAA are discussed. Results show an effective reduction of intensity fading for turbulent conditions that are even stronger than those expected in a natural environment. Experiments at 10 Gbit/s show the recovery of signal quality even in the presence of strong turbulence. In the current device, the power coupling efficiency is limited by the geometrical loss of the OAA, which is due to the separation of the GCs of the array that results in a fill factor of 4.3\% and a coupling loss of 13.6 dB. To improve the coupling efficiency, an array of microlenses can be integrated on top of the OOA. To this end, 3D printing techniques, such as two-photon polymerization (TPP), can be used to build custom-designed free-form optical elements directly on top of the photonic chip, thus facilitating optical alignment and reducing assembly and packaging costs \cite{LAM2022070025}.     

The beam contributions sampled by the OAA could also be added in the electrical domain. In this case, each optical aperture of the OAA has to be coupled to a fast PD, followed by a wide-band transimpedance amplifier and a receiver suitable for the specific signal data rate. The different electric signals can be combined by a multiple-input electronic digital signal processor, which has to operate at a bit-time scale. This solution, widely used in RF wireless spatial-diversity systems, becomes hardly practical, energy-intensive, very expensive when the data rate is raised at more than a few Gb/s, and sets strict constraints on the maximum number of apertures. Moreover, signal processing in the electrical domain is data- and protocol-dependent and unsuitable for simultaneous operation on multiple data channels transmitted in wavelength division multiplexed (WDM) optical systems. In our approach, the coherent combination of the signals in the optical domain employing an analog POP enables energy and cost-effective operation, scalability to a system with a larger number of apertures, modulation format transparency, extension to WDM and coherent optical signals, and pluggability with commercial fiber optics transceivers. From this point of view, the proposed multi-aperture receiver acts as an optical front-end interface to couple FSO beams (which are inherently multimode in the presence of turbulence) to single-mode guided optics. Once FSO signals are coupled to single-mode guiding structures (i.e., their spatial coherence is restored), they can be processed on a chip by using the portfolio of standard silicon photonics devices, and they can be efficiently coupled to fiber optics devices or even re-transmitted into free space link. \\

The proposed device concept can be ported to other wavelength ranges that are being explored for FSO communications, such as the mid infrared \cite{Didier:23} and the long wavelength infrared \cite{Joharifar_2023}, provided that suitable photonic integrated plaftorms enabling the realization of complex adaptive photonic integrated circuits are available. Moreover, the all-optical analog processing performed by our chip is expected to find applications also in quantum-communication FSO links, for instance, in terrestrial and space-to-earth quantum-key distribution systems \cite{Avesani_2021}. Finally, future works will also target the use of the POP-assisted OAA at the transmitter side of an FSO link to steer and optimally shape the optical beam to compensate for both turbulence and beam wander.

\section*{Methods}

\textbf{Numerical simulation of turbulence-induced scintillation}. The effects of atmospheric turbulence on arbitrary propagating beams are numerically evaluated by using a conventional split-step method \cite{Chatterjee} that considers sections of free space propagation regions (where only beam divergence occurs) alternated to phase screens (PSs). The PSs are generated using turbulence models, such as Kolmogorov, Von Karman, or modified Von Karman \cite{Andrews}, and the required number of PSs depends on the turbulence regime and on the accumulated scintillation along the path \cite{Zhan:20}. In this work, we considered a reference link distance of 800 m and turbulence regimes characterized by a refractive index parameter $C_n^2$ varying from $10^{-15}$ to $10^{-13} [\text{m}^{-2/3}]$. In a weak turbulence regime, the scintillation index equals about 0.1, and a single PS is sufficient. Instead, for the strongest turbulence regime considered, up to three PS should be used to guarantee the validity and accuracy of the split-step method.
\\

\noindent \textbf{Control electronics.} The control strategy for the automated self-configuration of the POP exploits local feedback loops to monitor and stabilize each MZI individually \cite{Miller:13}. To this end, dithering signals are applied to the thermo-optic phase shifters, thereby identifying deviations from the optimal bias point and maximizing the optical power detected at the PDs integrated at each MZI's output port\cite{Zanetto:1}. The PDs have a sensitivity of -40 dBm, thus enabling the control system to initiate automated configuration from arbitrary initial conditions of the POP and for any field pattern of the optical beam impinging on the OAA. Monitoring each MZI with a dedicated PD enables using a single dithering frequency for many actuators, allowing a wider control bandwidth. 
Alternatively, a single PD at the output port of the POP could be used, though this requires more complex electronics to read different pairs or dithering frequencies applied to the MZIs phase shifters \cite{seyedinnavadeh2024determining}. The electrical bandwidth of the electronic feedback controller (hosting 16 parallel lock-in chains to manage the 30 thermal phase shifters of the POP) is 2 kHz and enables a stabilization time of approximately 0.5 ms per MZI. The digital control electronics operate at 12 bits, ensuring a control voltage accuracy of about 1 mV under worst conditions (5V), corresponding to a phase shift uncertainty of approximately 1 mrad. In order to provide scalability to integrated POP-assisted receivers with more apertures, control electronics can be integrated into a multi-channel custom-designed CMOS application-specific integrated circuit (ASIC) that can automatically stabilize and reconfigure the MZI mesh.\cite{Sacchi:24}\\ 

\noindent \textbf{Experimental setup.}  With reference to Fig.~\ref{fig:SLM setup}a, a free-space Gaussian beam at a wavelength of 1550 nm is transmitted by a fiber-coupled optical collimator with a diameter of 3 mm. Controllable turbulence is intentionally introduced in the free-space link by using a spatial light modulator (SLM), which distorts the phase front of the beam by introducing a phase screen (PS) in the propagation path of the beam. Synthetic PSs were numerically generated to emulate a turbulence strength equivalent to several hundred meters of propagation with $C_n^2$ in the range $10^{-13}$ to $10^{-15} [\text{m}^{-2/3}]$ and then loaded in the SLM. A beam splitter was used to deviate a portion of the optical power to a near-infrared (IR) camera to capture the effects of the SLM on the beam. A 45°-tilted dielectric mirror and a 2-inch biconcave lens were used to couple the incoming beam to the OPA.  
\\
For dynamic tests, the SLM was replaced by a heat gun, and a beam expander was inserted at the transmitter side after the collimator. The beam expander generates a beam with a diameter of around 5 cm, and the heat gun is used to emulate the turbulence. By observing the beam scintillation of the IR camera, we estimated an effective refractive index perturbation up to $C_n^2$ of the order of $10^{-12}  [\text{m}^{-2/3}]$. The spectral components of the generated turbulence are shown in Supplementary Fig.~\ref{fig:sup turb FFT}. Due to the large size of the beam compared to the short length of the link (4 m), wander effects on the beam are negligible, and intensity fading is only due to scintillation. System-level measurements were performed by intensity modulating the transmitted beam at a data rate of 5 and 10 Gbit/s on-off keying (OOK) non-return-to-zero (NRZ). The wavelength for the 5 Gbit/s experiment was 1545 nm, whereas for the 10 Gbit/s was 1550 nm.

\bibliography{sample}

\section*{Acknowledgments}
The research has been carried out in the framework of the Huawei-Politecnico di Milano Joint Research Lab. Part of this work was carried out at Polifab (https://www.polifab.polimi.it), Politecnico di Milano, Milan, Italy. The Authors thank Photonpath srl (https://www.photon-path.com) for the chip-to-fiber transposer assembly.

\section*{Author contributions statement}
A.I.M. and G.C. developed the FSO simulator and performed the numerical analysis. S.S. designed the layout of the PIC. A.I.M. and S.S. designed and set up the experiment. A.I.M. carried out the experiments. A.I.M., A.M., and F.M. analyzed the results. F.Z. and M.S. supported the realization of the electronic control system. A. D. provided the industrial vision. A.M. and F.M. supervised the project. A.I.M., A.M., and F.M. wrote the manuscript. All the authors contributed to the revision of the manuscript.

\section*{Competing interests}
The authors declare no competing interests.

\section*{Additional information}
Correspondence and requests for materials should be addressed to A.M.









\pagebreak
\newpage

\section{Supplementary Information}
\subsection{Single-aperture receiver}
\label{sec: sup single aper}
Figure~\ref{fig:sup single aperture} displays the results for a simulation considering a Gaussian beam at the transmitter side with a beam waist $w_0=1$ mm that propagates for a distance  $L=800$ m. The single mode receiver is modeled as a Gaussian beam with $w_0=d_R/2$. The total power coupled to the receiver $P_{rx}$ is calculated based on the coupling coefficient $\eta$ as:
\begin{equation}
    \eta=\frac{|\int E^*_{rx}E_{tx} \,dA|^2}{\int |E_{tx}|^2\,dA \int |E_{rx}|^2\,dA}
\end{equation}
\begin{equation}
    P_{rx}= 10\log_{10}(\eta)  [\text{dB}]
\end{equation}
where $E_{rx}$ and $E_{tx}$ are the complex electric fields of the receiver and transmitted beam at the receiver plane. 
Panel (a) shows a simulation of the relative mean received power as a function of $d_R$ for no turbulence (purple), weak turbulence strength $C_n^2=10^{-15} [\text{m}^{-2/3}]$ (orange), moderate $C_n^2=10^{-14} [\text{m}^{-2/3}]$ (blue), and strong $C_n^2=10^{-13} [\text{m}^{-2/3}]$ (yellow). The corresponding values of $r_0$ are reported as reference.

\renewcommand{\thefigure}{S1}
\begin{figure}[b!]
\centering\includegraphics[width=17cm]{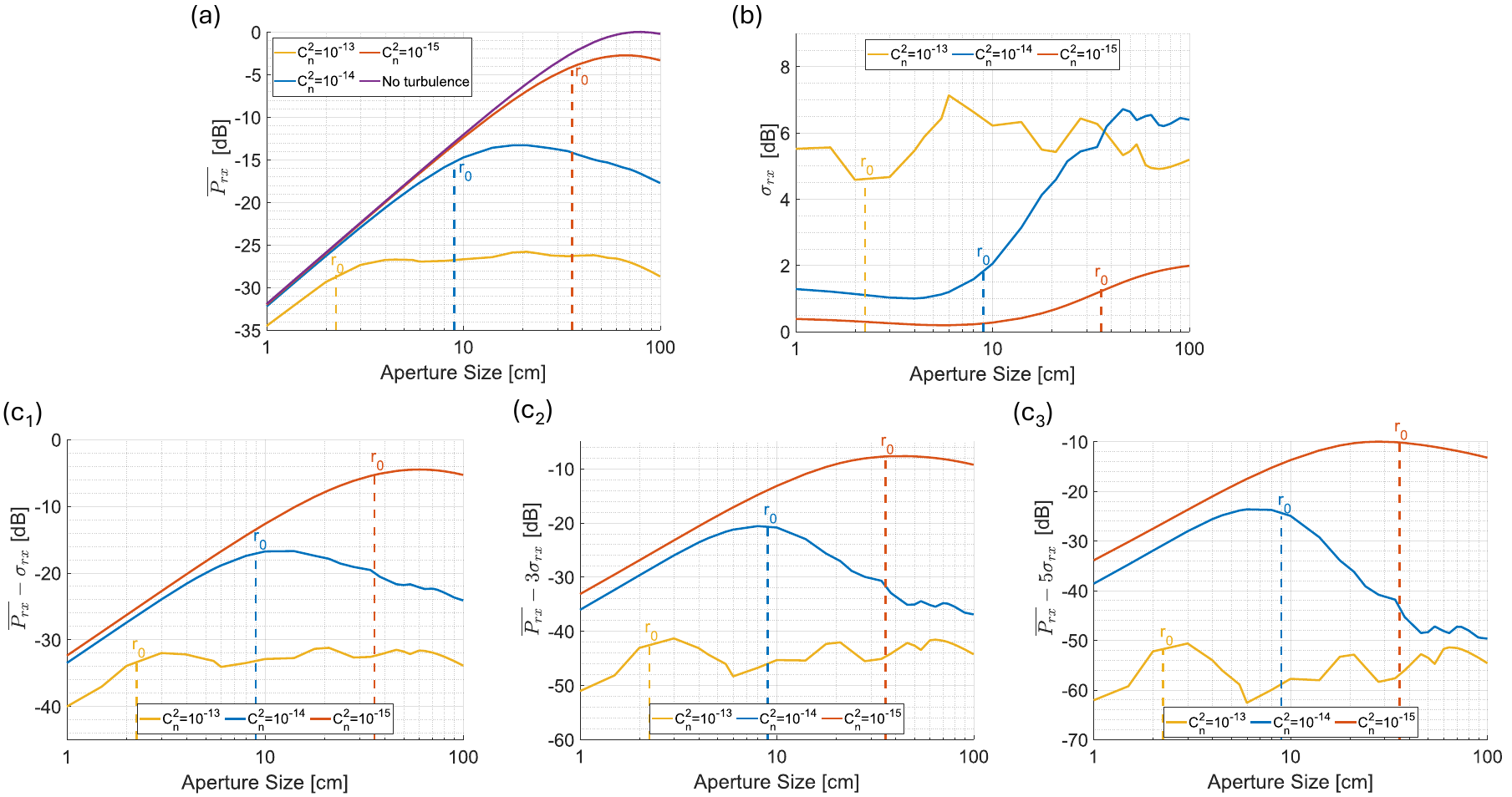}
\caption{\textbf{Single aperture receiver simulation}. \textbf{a-c}, Simulated mean received power (\textbf{a}), standard deviation (\textbf{b}), one-sigma (\textbf{c$_1$}), three-sigma (\textbf{c$_2$}), and five-sigma (\textbf{c$_3$}) limit of a single aperture receiver with single mode detector as a function of the aperture diameter $d_R$ for $C_n^2 =10^{-13}$ (blue), $10^{-14}$ (orange), $10^{-15}$ (yellow) $[\text{m}^{-2/3}]$, and without turbulence (purple).}
\label{fig:sup single aperture}
\end{figure}

A beam with an initial waist $w_0=1$ mm and wavelength $\lambda=1550$ nm has a Rayleigh length $z_R$: 
\begin{equation}
    z_R = \frac{\pi w_0^2}{\lambda}=2.0268  [\text{m}]
\end{equation}
 then, at $L=800$ m, the transmitted beam has a radius $w(z)$:
\begin{equation}
    w(z=800) = w_0 \sqrt{1+(\frac{z}{z_R})^2} = 39.5 [\text{cm}].
\end{equation}
Therefore, a receiver with $d_R=2w(z=800)=79$ cm will couple all the incoming power ($P_{rx}=0$ dB) in the absence of turbulence, as seen in the purple line.
Figure~\ref{fig:sup single aperture}b shows the corresponding standard deviation of $P_{rx}$, i.e., the fading. As a general rule, the stronger the turbulence, the lower the mean received power and the higher the fading.
In an FSO communication system, it is important to guarantee a low probability that the received power decreases below a given sensitivity threshold, say -18 dB, as an example. Figures~\ref{fig:sup single aperture}c show the (c$_1$) one-sigma, (c$_2$) three-sigma, and (c$_3$) five-sigma limit.

\subsection{Multi-aperture}
\label{sec: sup multi aper}
In section 3, we fixed $d_R = 4.5$ cm as a good compromise between a high $P_{rx}$ and low standard deviation $\sigma_{rx}$ for $C_n^2 \geq 10^{-13}$. The motivations behind this decision are displayed in Fig.~\ref{fig:sup multi aper size}. In these graphs, we plot the (a$_{1-3}$) mean received power and (b$_{1-3}$) standard deviation as a function of the separation distance $d_S$ for an aperture size $d_R = 2$ (blue), 4.5 (orange), and 18 (yellow) cm; a number of antennas $M=1$ (dashed line), 2 (circles), 5 (diamonds), and 16 (squares); and a turbulence parameter $C_n^2= 10^{-15}$ (column 1), $10^{-14}$ (column 2), and $10^{-13}$ (column 3) $[\text{m}^{-2/3}]$. As can be seen, the largest aperture couples more power for medium scintillation but underperforms for strong scintillation. In fact, the OAA with $M=16$ and $d_R=4.5$ cm is the one that couples more power for $C_n^2=10^{-13}$ $[\text{m}^{-2/3}]$. Also, as in the case of a single aperture, the $\sigma_{rx}$ increases as the aperture size does; therefore, a large antenna is not recommended. When comparing $d_R=2$ and $d_R=4.5$ cm, the latest is the best choice given the higher $P_{rx}$ and equal or lower $\sigma_{rx}$.

\renewcommand{\thefigure}{S2}
\begin{figure}[h!]
\centering\includegraphics[width=17cm]{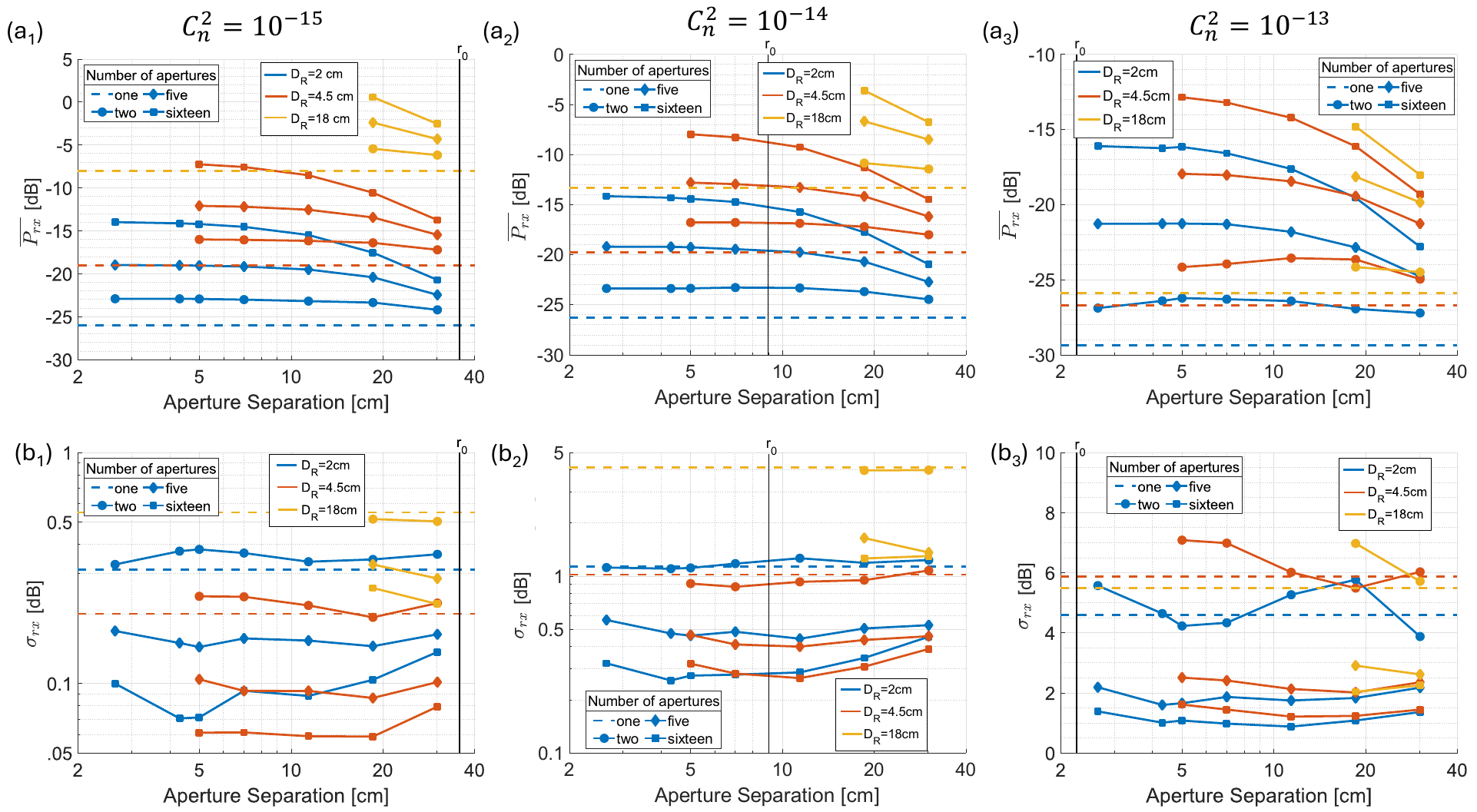}
\caption{\textbf{Multi-aperture receiver simulation for different aperture sizes}. Each column corresponds to one of the considered turbulences: $C_n^2= 10^{-15}$, $10^{-14}$, and $10^{-13}$ $[\text{m}^{-2/3}]$. \textbf{a-b}, Mean received power (\textbf{a}) and standard deviation (\textbf{b}) as a function of the separation distance $d_S$ for an aperture size $d_R = 2$ (blue), 4.5 (orange), and 18 (yellow) cm. The number of antennas is $M=1$ (dashed line), 2 (circles), 5 (diamonds), and 16 (squares). $P_R$ is calculated by summing coherently the contributions of the $M$ apertures.}
\label{fig:sup multi aper size}
\end{figure}

Figure~\ref{fig:sup multi aper} shows the same results as Fig.~\ref{fig:multi aperture}, that is, (a) mean received power, (b) standard deviation, and (c) three-sigma but not as a function of the number of antennas, but the separation distance $d_S$. The larger the separation, the antennas move away from the centroid of the beam, and the lower the power coupled to the receiver. The $\sigma_{rx}$ is almost constant for all $d_s$, except when it is too large, which starts to increase.   

\renewcommand{\thefigure}{S3}
\begin{figure}[ht!]
\centering\includegraphics[width=17cm]{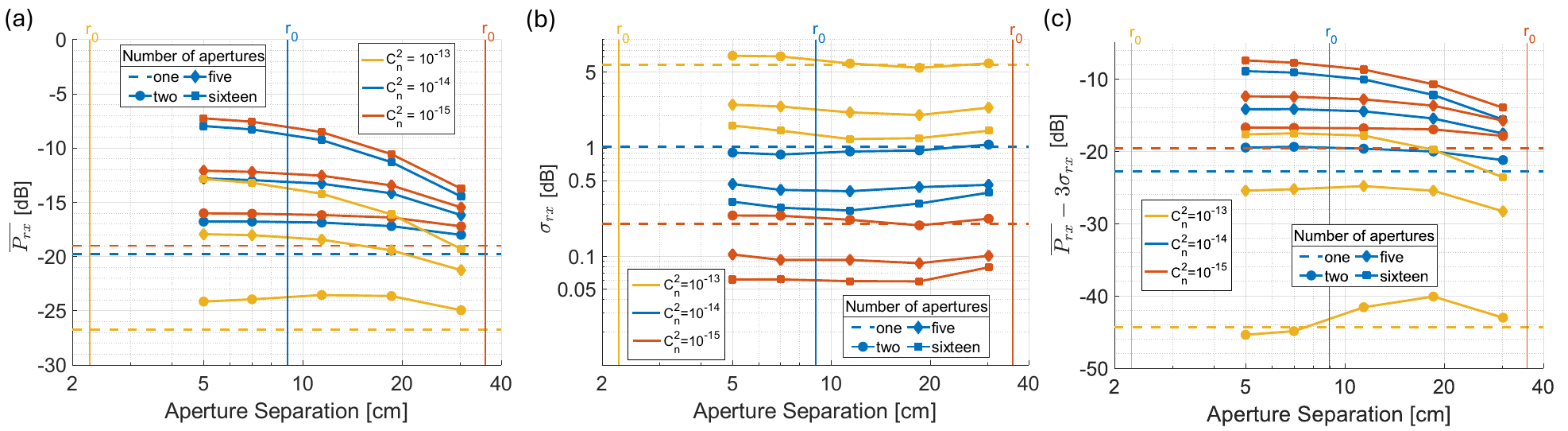}
\caption{\textbf{Multi-aperture receiver simulation for different aperture separations}. \textbf{a-b}, Mean received power (\textbf{a}), standard deviation (\textbf{b}), and three-sigma (\textbf{c}) as a function of the separation distance $d_S$ for a number $M=1$ (dashed line), 2 (circles), 5 (diamonds), and 16 (squares) of apertures and the three considered turbulences $C_n^2= 10^{-13}$ (yellow), $10^{-14}$ (orange), and $10^{-15}$ (blue) $[\text{m}^{-2/3}]$. $d_R = 4.5$ cm. $P_R$ is calculated by summing coherently the contributions of the $M$ apertures.}
\label{fig:sup multi aper}
\end{figure}

\newpage
\textbf{}
\newpage

\section{Supplementary Figures}
\vspace{2cm} 
\renewcommand{\thefigure}{S4}
\begin{figure}[ht!]
\centering\includegraphics[width=17cm]{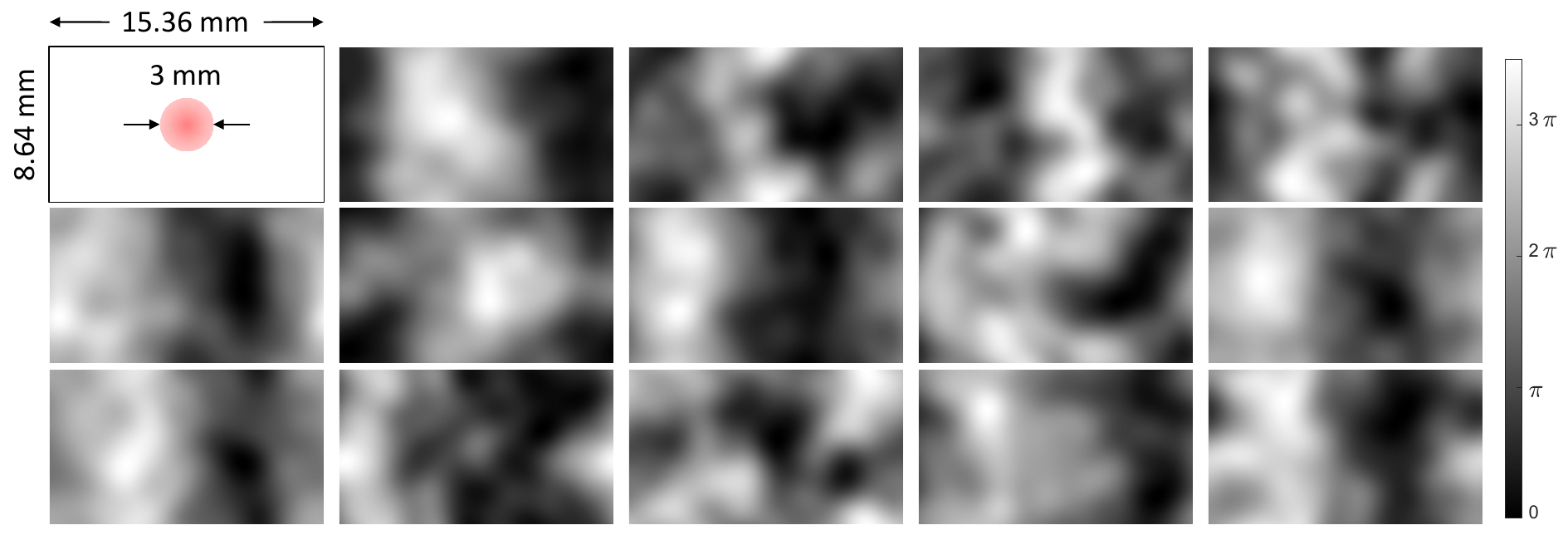}
\caption{\textbf{Phase screens used to emulate the turbulence with the SLM}. See  Fig.~\ref{fig:SLM setup}d. The red spot in the first PS, illustrates the size of the beam.}
\label{fig:sup SLM ps}
\end{figure}
\vspace{2cm} 
\renewcommand{\thefigure}{S5}
\begin{figure}[ht!]
\centering\includegraphics[width=8cm]{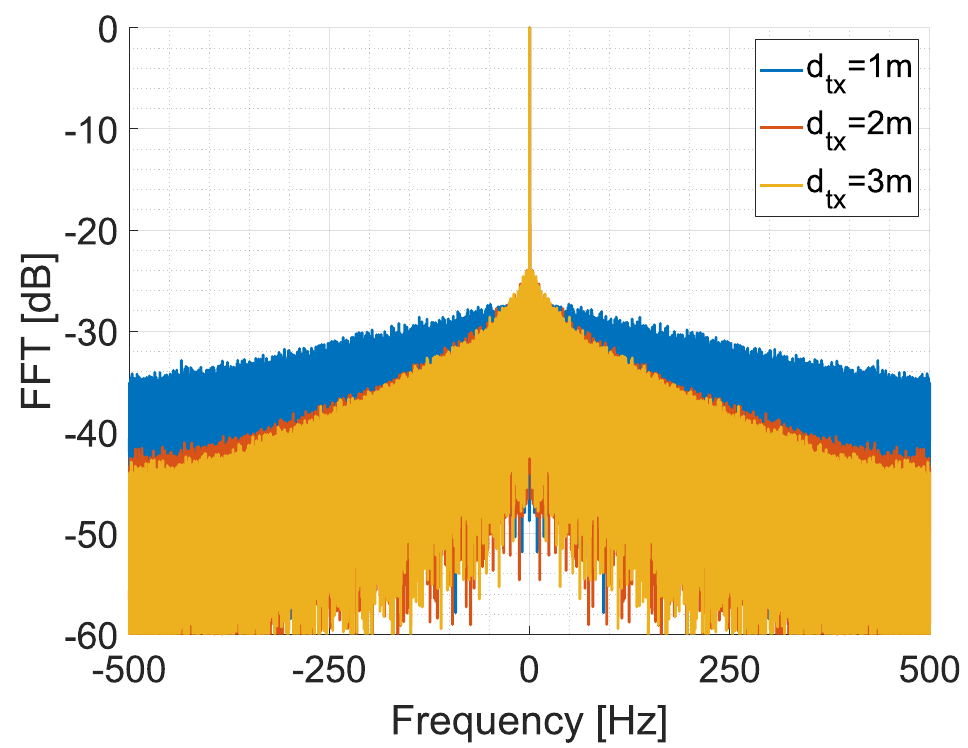}
\caption{\textbf{Emulated turbulence spectrum}. FFT of the time trace acquired with a 3 mm collimator for 3 distances of the thermal gun from the transmitter $d_{tx}=1$, 2, and 3 m. The closer the gun is to the transmitter, the stronger the turbulence.}
\label{fig:sup turb FFT}
\end{figure}

\end{document}